\documentclass{emulateapj}
\usepackage{wasysym}
\usepackage{apjfonts}
\usepackage{graphicx}
\usepackage{amssymb}
\usepackage{dsfont}
\usepackage{textcomp}

\renewcommand{\earth}{\oplus}

   \submitted{Accepted for publication in The Astrophysical Journal}

\newcommand{\tdv}{T$\delta$V$\,$}
\newcommand{\di}{\Delta i}

\begin{document}

\bibliographystyle{apj}

\title{The Detectability of Transit Depth Variations due to \\
  Exoplanetary Oblateness and Spin Precession }

\author{Joshua A.~Carter \& Joshua N.~Winn}

\affil{Kavli Institute for Astrophysics and Space Research,
  \\Massachusetts Institute of Technology, Cambridge, MA 02139 \\
  carterja@mit.edu}

\begin{abstract}

  Knowledge of an exoplanet's oblateness and obliquity would give
  clues about its formation and internal structure. In principle, a
  light curve of a transiting planet bears information about the
  planet's shape, but previous work has shown that the
  oblateness-induced signal will be extremely difficult to
  detect. Here we investigate the potentially larger signals due to
  planetary spin precession. The most readily detectable effects are
  transit depth variations (\tdv) in a sequence of light curves. For a
  planet as oblate as Jupiter or Saturn, the transit depth will
  undergo fractional variations of order $1\%$. The most promising
  systems are those with orbital periods of approximately 15--30 days,
  which is short enough for the precession period to be less than
  about 40 years, and long enough to avoid spin-down due to tidal
  friction. The detectability of the T$\delta$V signal would be
  enhanced by moons (which would decrease the precession period) or
  planetary rings (which would increase the amplitude). The {\em
    Kepler} mission should find several planets for which
  precession-induced T$\delta$V signals will be detectable.  Due to
  modeling degeneracies, {\em Kepler} photometry would yield only a
  lower bound on oblateness.  The degeneracy could be lifted by
  observing the oblateness-induced asymmetry in at least one transit
  light curve, or by making assumptions about the planetary interior.

\end{abstract}


\maketitle

\section{Introduction}

Measuring the oblateness of exoplanets would further our understanding
of planetary formation, rotation, and internal structure. One possible
measurement technique relies on the differences between the transit
light curve of a spherical planet and an oblate planet with the same
sky-projected area (Seager \& Hui 2002, Barnes \& Fortney
2003). However, the differences are minuscule, of order 200 parts per
million (ppm) for a planet as oblate as Saturn, and 2~ppm for a ``hot
Jupiter'' whose spin rate has been slowed by tidal friction into
synchronization with its orbit. In a previous paper we showed that
with this technique, even the best available light curves are only
barely sufficient to rule out a Saturn-like oblateness (Carter \& Winn
2009). Those results pertained to the planet HD~189733b, for which one
would expect spin-orbit synchronization, and consequently the
theoretical oblateness was an order of magnitude below the empirical
upper limit.

In that work we also pointed out the potentially observable effects of
a phenomenon that had been previously overlooked: the precession of
the planet's rotation axis. Precession of an oblate planet causes the
sky-projected area of the planet to change over time, thereby causing
gradual changes in the depth and duration of transits. In this paper
we investigate the observable manifestations of spin precession in a
broader context. Section 2 describes the characteristics of the signal
in terms of the properties of the star, planet, and orbit. Section 3
discusses the range of orbital periods for which the signal is most
readily detectable, considering the timescales for precession and
spin-orbit synchronization. Section 4 presents simulated results for a
specific case, a Saturn-like planet observed by the {\it Kepler}
satellite. Section 5 summarizes and discusses the results.
\\
\\

\section{Characteristics of the signal} \label{sec:proj}

\begin{figure*}[htbp] 
   \centering
   \epsscale{0.8}
     \plotone{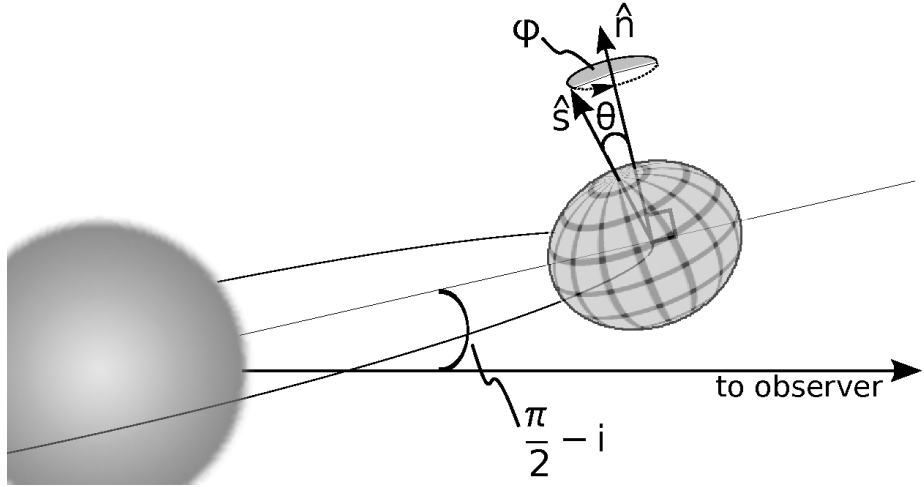}
     \caption{Geometry of transits by an oblate spheroid. The planet's
       orientation is specified by the spin-orbit obliquity $\theta
       \equiv \cos^{-1} \hat{s}\cdot\hat{n}$, the angle $\phi$ between
       the projection of the spin axis onto the orbital plane and the
       vector connecting the centers of the star and planet at
       midtransit, and the orbital inclination $i$ relative to the sky
       plane. }
   \label{fig:dia}
\end{figure*}

We model the planet as an oblate spheroid, illustrated in
Figure~(\ref{fig:dia}). The {\it oblateness} (or {\it flatness})
parameter is defined as
\begin{equation}
 f = \frac{R_{\rm eq} - R_{\rm pol}}{R_{\rm eq}},
 \label{eq:oblateness-definition}
\end{equation}
where $R_{\rm eq}$ and $R_{\rm pol}$ are the equatorial and polar
radii, respectively. For rotationally-induced oblateness, a good approximation is
\begin{eqnarray}
 f = \frac{3}{2} J_2 +
 \frac{1}{2}\frac{R_{\rm eq}^3}{G M_p} \left(\frac{2 \pi}{P_{\rm rot}}\right)^2
 \label{eq:rot}
\end{eqnarray}
where $M_p$ is the mass of the planet and $J_2$ is the planet's zonal
quadrupole moment (Murray \& Dermott 2000, Hubbard 1984). This
approximation is valid for all the Solar System planets for which $f$,
$J_2$ and $P_{\rm rot}$ have been measured precisely.

The {\it obliquity} is the angle $\theta$ between the polar axis and
the orbital axis. The angle $\phi$ specifies the direction of the
projection of the polar axis onto the orbital plane. For a uniformly
precessing planet, the case considered here, $\phi$ is a linear
function of time,
\begin{equation}
\phi(t) = \frac{2\pi t}{P_{\rm prec}} + \phi_0.
\end{equation}

Accurate calculations of the transit light curve of an oblate
spheroid, including the effects of stellar limb darkening, have been
presented by Seager \& Hui~(2002), Barnes \& Fortney~(2003), and
Carter \& Winn~(2009). In this paper we are not interested in high
accuracy or in the slight differences between the light curve an
oblate planet and a spherical planet. Instead we are interested in the
order of magnitude of the variations in the transit depth and duration
due to the changes in the precessing planet's sky-projected figure.

In the absence of limb darkening, the depth $\delta$ is approximately
the areal ratio between the sky projection of the oblate spheroid and
the stellar disk,
\begin{eqnarray}
\delta(t) & = & k^2 \sqrt{1 - \epsilon^2 \left\{1-\left[\sin \theta\cos\phi(t) \sin i + \cos\theta \cos i\right]^2\right\}}
\label{eq:depth}
\end{eqnarray}
where $k \equiv R_{\rm eq}/R_\star$ is the planet-to-star radius
ratio, $i$ is the orbital inclination with respect to
the sky plane and $\epsilon$ is the {\it ellipticity},
\begin{equation}
\epsilon \equiv \sqrt{1-(1-f)^2}.
\label{eq:ellipticity}
\end{equation}
A derivation of this expression is given in the Appendix.  Figure~\ref{fig:amps} shows the fractional amplitude of the transit
depth variations for the case $i = 90^\circ$, as a function of $f$ and
$\theta$. For Saturn-like values of oblateness and obliquity, the
depth variations would be a few percent.

\begin{figure*}[htbp] 
   \centering
   \epsscale{0.94}
     \plotone{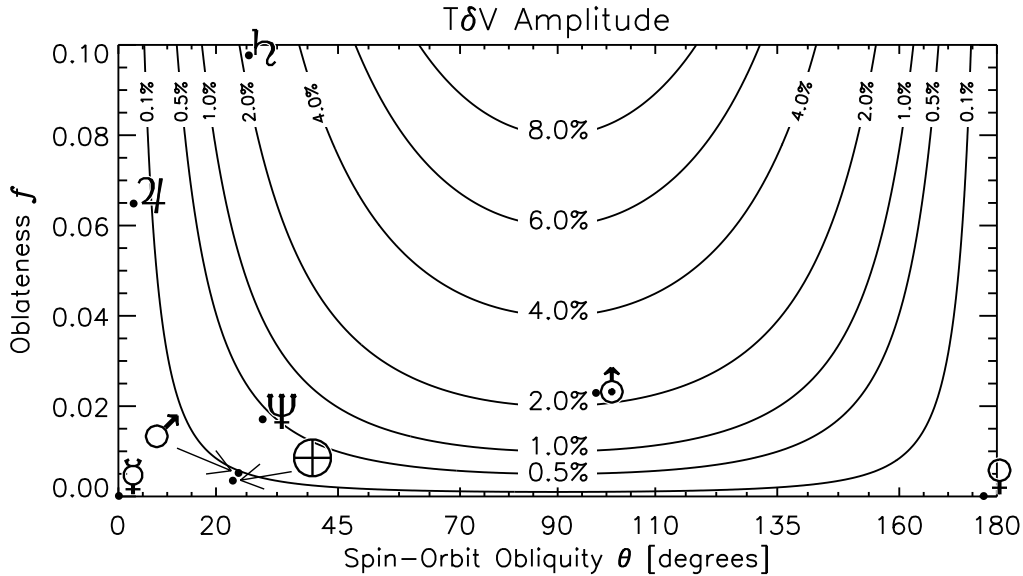}
     \caption{The fractional amplitude of transit depth variations
       (T$\delta$Vs), shown as a function of oblateness ($f$) and
       obliquity ($\theta$) assuming the orbit is perpendicular to the
       sky plane ($i = 90^\circ$). For reference, the astrological
       symbols show the shape parameters of Solar System planets:
       $\mercury$ -- Mercury, $\venus$ -- Venus, {$\earth$} -- Earth,
       $\mars$ -- Mars, $\jupiter$ -- Jupiter, $\saturn$ -- Saturn,
       $\uranus$ -- Uranus, $\neptune$ -- Neptune.}
   \label{fig:amps}
\end{figure*}

The transit duration will also vary over the precession period, due to
the changing dimension $R_\parallel$ of the planet's sky projection in
the direction of orbital motion:
\begin{equation}
  R_\parallel = R_{\rm eq} \sqrt{\left(1-f_\perp\right)^2 \sin^2 \theta_\perp + \cos^2 \theta_\perp},
\end{equation}
where
\begin{eqnarray}
	f_\perp & \equiv & 1- \sqrt{1 - \epsilon^2 \left\{1-\left[\sin \theta\cos\phi(t) \sin i + \cos\theta \cos i\right]^2\right\}} \label{eq:fperp}\\
	\theta_\perp & \equiv & \tan^{-1} \frac{\sin \phi(t) \tan \theta}{\cos \phi(t) \tan \theta \cos i - \sin i} \label{eq:thetaperp}
\end{eqnarray}
are the quantities describing the oblateness and obliquity of the
projected exoplanet's shape.  Derivations of these expressions are also given in the Appendix. For a circular orbit, the ingress/egress
duration (first to second contact, or third to fourth contact) is
approximately
\begin{eqnarray}
	\tau \approx  \left(\frac{R_\star P_{\rm orb}}{\pi a}\right)
        \frac{R_\parallel}{R_\star}\sqrt{1-b^2},
\end{eqnarray}
where $P_{\rm orb}$ is the orbital period, $a$ is the orbital
distance, and $b \equiv a \cos i/ R_\star$ is the normalized impact
parameter. The full transit duration (first to fourth contact) is
approximately (e.g., Seager \& Mall{\'e}n-Ornelas 2003)
\begin{equation}
  T_{\rm full} \approx \left(\frac{R_\star P_{\rm orb}}{\pi a} \right)
     \sqrt{ \left[1 + \frac{R_\parallel}{R_\star} \right]^2 - b^2 }. \label{eq:dur}
\end{equation}
These approximations are valid as long as the transit is not too close
to grazing. The fractional amplitude of the $\tau$ variations
(T$\tau$V) is comparable to that of depth variations (T$\delta$V).
The amplitude of transit full-duration variations (TDV) depends on $k$
and $b$, in addition to $f$ and $\theta$, and therefore cannot be
summarized in a single contour plot such as Figure~\ref{fig:amps}.

\section{The optimal orbital distance}
\label{sec:times}

In this section we suppose that the planet's precession is caused
exclusively by the gravitational torque from the star. In order for
the detection of precession-induced T$\delta$Vs or TDVs to be
feasible, the planet must be close enough to the star for precession
to produce observable effects in a human lifetime. However, if the
planet is too close to the star, then tidal dissipation should slow
down the planet's rotation until it is synchronized with the orbital
period, and drive the obliquity to zero, which would cause the signal
to be undetectable. Hence, we must ask if there is a range of
distances from the star that is close enough for rapid precession, and
yet far enough to avoid spin-orbit synchronization.

\begin{figure}[h] 
   \centering
  \epsscale{1.3}
   \hspace*{-0.2in}\plotone{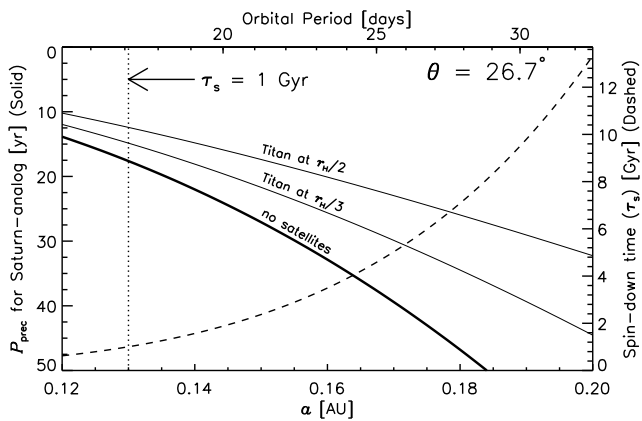}
\caption{
The optimal orbital distance. The solid curves refer to the axis on
the left; they show the calculated spin precession period as a
function of orbital distance for a Saturn-analog planet, with no
satellites (thick line) and with a Titan-like satellite (thin lines)
orbiting at a specified fraction of a Hill radius. The dashed curve
refers to the axis on the right; it shows the calculated timescale for
tidal spin-orbit synchronization. The vertical dotted line marks the orbital distance
for which the spin-down time is 1 Gyr.
\label{fig:times}}
\end{figure}

The spin precession period for a planet on a fixed circular orbit is
given by
\begin{eqnarray}
P_{\rm prec} = \frac{13.3~{\rm yr}}{\cos\theta}
               \left(\frac{\mathds{C}/J_2}{13.5}\right)
               \left(\frac{P_{\rm orb}}{15~{\rm d}}\right)^2
               \left(\frac{10~{\rm hr}}{P_{\rm rot}}\right)
  \label{eq:prec}
\end{eqnarray}
(Ward 1975), where $P_{\rm rot}$ is the planet's rotation period and
$\mathds{C}$ is its moment of inertia divided by $M_p R_{\rm
  eq}^2$. The numerical scaling of 13.5 for $\mathds{C}/J_2$ is the
estimated value for Saturn (Ward \& Hamilton 2004). According to this
expression, orbital periods shorter than $P_{\rm orb}\sim$30~days will lead to
rapid enough precession to be observed over decadal timescales,
depending on the planet's obliquity and internal structure.  In
Fig.~\ref{fig:times}, the thick solid line shows the spin precession
period as a function of orbital distance, for an exoplanet with the
same $P_{\rm rot}$, $\mathds{C}$, $J_2$, and $\theta$ as Saturn. (The
thin solid lines show the more rapid precession rates produced by
hypothetical planetary satellites, as discussed in \S~\ref{sec:disc}.)

The approximate timescale for tidal spin-orbit synchronization is
\begin{eqnarray}
  \tau_{\rm s} \approx 1.22~{\rm Gyr} &\times&
        \left(\frac{Q_p}{10^{6.5}}\right)
        \left(\frac{\mathds{C}}{0.25}\right)
        \left(\frac{5~{\rm hr}}{P_{{\rm rot},i}}-\frac{10~{\rm hr}}{P_{\rm rot}}\right) \nonumber \\
          &\times& 
        \left(\frac{M_p}{M_{\rm Jup}}\right) \left(\frac{R_{\rm Jup}}{R_{\rm eq}}\right)^3
        \left(\frac{P_{\rm orb}}{15~{\rm d}}\right)^4
   \label{eq:tsync}
\end{eqnarray}
(Goldreich \& Soter 1966), where $Q_p$ is the planet's tidal
dissipation factor and $P_{{\rm rot},i}$ is its initial rotation
period, both of which are highly uncertain.
 
Based upon Solar System constraints, $Q_p$ is thought to be in the
range 10--500 for terrestrial planets, and $10^{4.5\mbox{--}6.5}$ for
gas or ice giant planets (Goldreich \& Soter 1966, Peale et al.~1980,
Yoder 1995, Mardling \& Lin 2004, Ogilvie \& Lin 2004, Jackson et
al.~2008), although the results are strongly model-dependent and also
dependent on the frequency of tidal oscillations. In setting the scale
parameters in Eqn.~(\ref{eq:tsync}) we adopted $Q_p=10^{6.5}$, on the
high end of current estimates, giving the longest (most favorable)
synchronization timescale.

The primordial spin period might be expected to be near the rotational
breakup limit, which is $\approx$3~hr for Jupiter (Murray \& Dermott
2000), although the effects of planetary contraction and disk-planet
interactions should also be considered. For Eqn.~(\ref{eq:tsync}) we
used a current rotation period of $10$~hr (similar to Jupiter and
Saturn) and an initial period of $5$~hr.

For $\tau_{\rm s}\gtrsim 1$~Gyr, it is reasonable to hope that the
planet has not yet been tidally spun down. Thus, with reference to
Eqns.~(\ref{eq:prec}) and (\ref{eq:tsync}), the ``sweet spot'' for
observing the effects of spin precession on the transit parameters is
at $P_{\rm orb}\sim 15$~d, which is long enough to allow for rapid
rotation, and short enough to allow for rapid precession.

Of course the identification of a single optimal period is a
simplification. The existence and observability of T$\delta$Vs and
TDVs depends on the particular mass and radius of the planet under
consideration and the observed age of the star, as well as the
parameters relating to the planet's tidal dissipation, internal
constitution, and initial spin period. The dashed line in
Figure~\ref{fig:times} shows the dependence of $\tau_{\rm s}$ on
orbital distance for the particular case of a Saturn
analog. Table~\ref{tab:planets} gives some numerical results for
precession periods and synchronization timescales for hypothetical
close-in planets with properties similar to Solar System planets. If
we require $P_{\rm prec}<40$~yr and $\tau_{\rm s}>1$~Gyr for
observability, then from Table~\ref{tab:planets} we see that the
signal is potentially observable for Jupiter and Saturn analogs. For
analogs of Uranus and Neptune, we expect the signal to be unobservable
unless the spin precession is made more rapid by the presence of large
satellites (see \S~\ref{sec:disc}). For Earthlike planets the signal
seems unlikely to be observable because of strong tidal dissipation.

\begin{deluxetable*}{lcccccccccccc}\vspace{-0.3in}
\tablecolumns{13}
\tablewidth{0pt}
\tablecaption{Relevant timescales for the observability of planetary spin precession}
\tablehead{& \multicolumn{5}{c}{{Adopted planetary parameters}}  &\hspace{0.001in}& \multicolumn{2}{c}{{for $\tau_{\rm s} = 1$ Gyr}} &\hspace{0.001in} &\multicolumn{2}{c}{{for $P_{\rm prec} = 40$ yr}} &\\ \cline{2-6} \cline{8-9} \cline{11-12} \vspace{-0.1in}\\
		{Planet analog\tablenotemark{a}} &{$P^i_{\rm rot}$ [hr]}&{$P_{\rm rot}$ [hr]} &{$Q_p$} &{$\mathds{C}$} &{$\mathds{C}/J_2$} & & $P_{\rm prec}$ [yr] & $P_{\rm orb}$ [d]  && $P_{\rm orb}$ [d] & $\tau_s$ [Gyr] & Observable Range in $P_{\rm orb}$ [d]} 
\startdata
Jupiter & 5.0 & 10.0 & $10^{6.5}$ & 0.26&15.4 & & 13.7 &{14.3 }&& {24.3} & 8.5 & 14.3 $-$ 24.3\nl
Saturn & 5.0 &11.0 & $10^{6.5}$ & 0.22& 13.5& &15.7& {17.1} && {27.3}\vspace{0.05in}& 6.5 & 17.1 $-$ 27.3\nl 
Uranus & 8.0 &17.0 & $10^{6.5}$ & 0.22& 67.0& & 45.3 & 16.3 && 15.3 & 0.78 & None\tablenotemark{a}\nl  
Neptune & 8.0 &16.0  & $10^{6.5}$ &  0.23& 68.0& & 43.8 & 15.3 && 14.7 & $0.83$ & None\tablenotemark{a}\vspace{0.05in}\nl 
Earth & 12.0 & 24.0 & $10^{2.0}$ & 0.34 & 314 & & 13,100 &151 && 8.40 & $9.0\times 10^{-6}$ &None\tablenotemark{a}\nl\vspace{-0.05in}
\enddata
\label{tab:planets}
\tablenotetext{a}{Assuming no satellites}
\tablerefs{Murray \& Dermott (2000), Hubbard (1984), Yoder (1995), Ward \& Hamilton (2004)}
\end{deluxetable*}

\section{A specific example}
\label{sec:case}

In this section we examine the particular case of a hypothetical,
ringless, moonless, Saturn-like transiting exoplanet ($f=0.1$, $\theta=27^\circ$) in a
circular orbit around a Sun-like star with $P_{\rm orb}=17$~d. This
orbit is near the ``sweet spot,'' giving $P_{\rm prec} = 17$~yr and
$\tau_{\rm s} = 1$~Gyr.

\subsection{Likelihood of discovery}

First we must ask how likely it is that such a planet will be
discovered. Giant planets with periods between 15--30~d are already
known to exist from Doppler surveys. At the time of writing, the {\tt
  exoplanets.org} database has 8 such planets. Indeed one of them is
already known to transit (HD~17156b; Fischer et al.~2007), although in
that case tidal effects may have slowed the planet's rotation because
the orbit is highly eccentric and the pericenter distance is
small. None of the other 7 planets is known to transit, and the
probability that at least one of them transits is approximately
23\%. Given these facts and the recent acceleration in the discovery
rate using the Doppler method, it would seem likely that a transiting
gas giant with $P_{\rm orb}=15$--30~d will be discovered in the near
future.

Of particular interest are the prospects for the {\it Kepler}
satellite mission (Borucki et al.~2010). {\it Kepler} searches for
transiting planets by keeping $\approx$10$^5$ stars under nearly
continuous photometric surveillance for at least 3.5~yr and possibly
as long as 6~yr. To estimate the fraction of the target stars that has
a suitable planet, we used the power-law formulas given by Cumming et
al.~(2008) for the abundance of planets of a given mass and period,
which were derived from data from the Keck Planet Search. The
integrated value of (abundance~$\times$~transit probability) over the
range $P_{\rm orb}=$15--30~d, $M_p=$~0.2--2~$M_{\rm Jup}$ is
approximately 0.019\%. Therefore, among the $5\times10^4$ Sun-like
stars in the {\it Kepler} field, we expect $\approx$10 transiting
giant planets with $P_{\rm orb} =$~15--30~d.

\subsection{Expected signal}

\begin{figure*}[t] 
   \centering
   \epsscale{0.95}
      \plotone{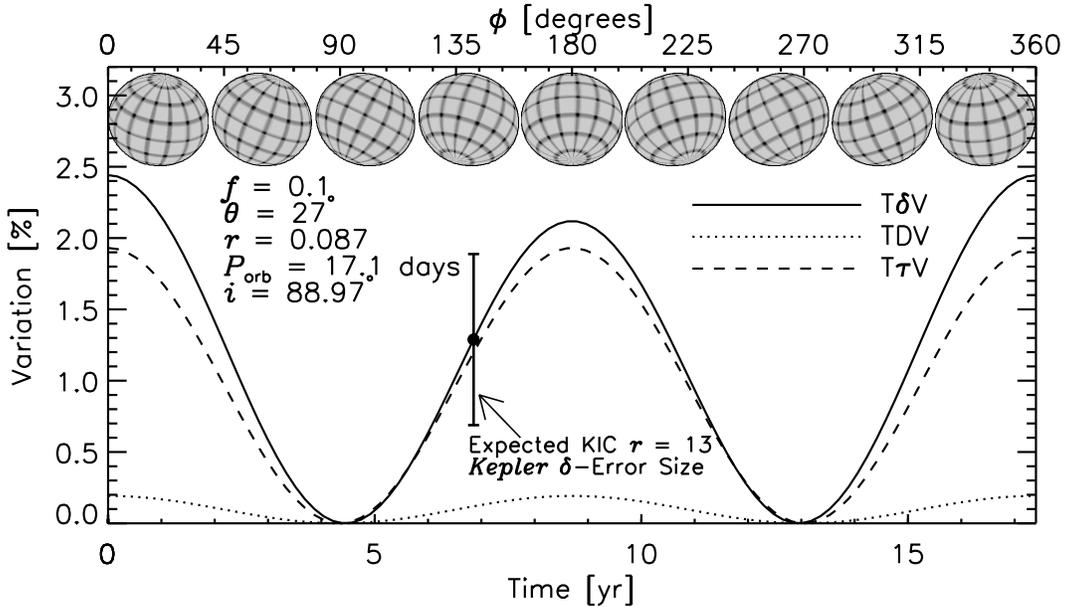}
      \caption{Variations in the transit light curve due to an oblate,
        oblique, precessing exoplanet. Plotted are the transit depth
        ($\delta$), total duration ($T_{\rm full}$) and ingress duration ($\tau$)
        fractional variations (\tdv, TDV, and T$\tau$V, respectively) that are expected for a uniformly precessing
        Saturn-like planet around a Sun-like star.  The time scale is
        based on the assumption $P_{\rm orb} = 17.1$~days.}
   \label{fig:curve}
\end{figure*}

Figure~\ref{fig:curve} shows the expected time variations in $\delta$,
$\tau$, and $T_{\rm full}$ for our hypothetical close-in
Saturn. Compared to the T$\delta$V signal, the TDV signal has a
smaller amplitude, and depends on a larger number of parameters.  The
T$\tau$V signal has a similar amplitude to the T$\delta$V signal, but
$\tau$ cannot be measured as precisely as $\delta$ because of the
relatively short duration of the partial transit phases. Therefore we
expect the T$\delta$V signal to provide the best constraints on the
planetary shape parameters.

Depth variations of a few percent should be detectable. Ground-based
observations of individual transits have already allowed $\delta$ to
be measured to within 1\% [see, e.g., Gillon et al.~(2009), Johnson et
al.~(2009), Winn et al.~(2009)]. To estimate the precision with which
{\it Kepler} can measure transit depth variations, we simulated a
transit light curve and averaged it into 30~min bins, corresponding to
{\it Kepler}'s standard time sampling.  Then we added white noise with
a standard deviation of $95$~ppm, as appropriate for a typical {\it
  Kepler} target (KIC $r$~mag~$=13$), based on the early results of
Welsh et al.~(2010) and Latham et al.~(2010). By fitting a
parameterized model to the light curve we found that the transit depth
was recovered to within $0.6\%$. Over a 6~yr mission, {\em Kepler}
would observe $\approx$130 transits with this precision.

\subsection{Parameters and degeneracies}

Next we discuss the ``inverse problem'' of inferring the planetary
shape and precession parameters from the observed T$\delta$V signal.
For $i=90^\circ$, the T$\delta$V signal is periodic with period
$P_{\rm prec}/2$. For $i\neq 90^\circ$ the T$\delta$V period is
actually $P_{\rm prec}$, but the odd and even maxima have only
slightly different amplitudes. Hence the period of the T$\delta$V
signal will reveal $P_{\rm prec}$.

Three other characteristics of the T$\delta$V signal will help to pin
down the model parameters. First, the minimum observed transit depth
will reveal the parameter combination
\begin{eqnarray}
\delta_{\rm min} & = & k^2 \left(1-f\right) = k^2 \sqrt{1-\epsilon^2}.
 \label{eq:quant1}
\end{eqnarray}
Second, the full range of transit depths determines a combination of
$f$ and $\theta$,
\begin{eqnarray}
 \left[
  \frac{ \delta_{\rm max} }{ \delta_{\rm min} }
 \right]^2-1 & = & \frac{\epsilon^2}{1-\epsilon^2}\left( \sin i \sin \theta+\cos i \cos \theta\right)^2\nonumber \\
 	& = & \frac{\epsilon^2}{1-\epsilon^2} \left(\sin^2 \theta+\di \sin 2\theta \right)+O(\di)^2  \label{eq:quant2}\\
 & \approx & 2 f \sin^2 \theta, \nonumber
\end{eqnarray}
where $\di = \pi/2-i$ is always small and can be measured precisely
through transit photometry. Third, the phases at which the maxima
occur will determine $\phi_0$.

For $\di \neq 0$ there are two other informative observables, at least
in principle. First, the phases at which the minima occur depends
slightly on $\di$ and $\theta$:
\begin{eqnarray}
\phi_{\rm min}-\phi_0 & = & (2n-1) \frac{\pi}{2}+\left\{ \begin{array}{cl}
						+\di \cot \theta  & \mbox{,~$n$ odd} \\
						-\di \cot \theta & \mbox{,~$n$ even}
								\end{array} \right. + O(\di)^3,
\end{eqnarray}
where $n = $1, 2, 3,~$....$ Second, there is a slight difference in
heights between odd and even maxima,
\begin{eqnarray}
\Delta \delta_{ \rm max}
 & = & \di \frac{k^2 \epsilon^2 \sin 2\theta}{\sqrt{1-\epsilon^2 \cos^2 \theta}}+O\left(\di\right)^3,
\end{eqnarray}
but these effects are very small in practice. For our Saturn analog,
assuming $i = 88.97^\circ$ (transit impact parameter~$=0.5$), the
difference in the maxima is smaller than 0.25\%.

\begin{figure}[h] 
   \centering
   \epsscale{1.1}
  \plotone{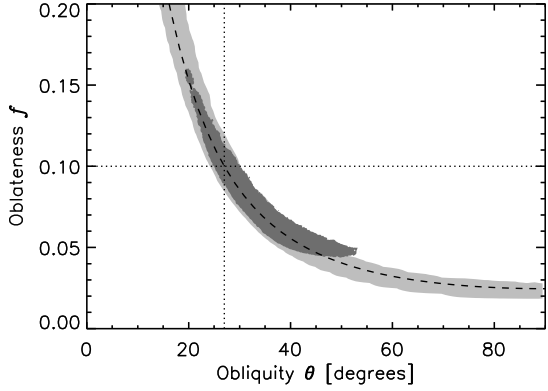}
  \caption{ Hypothetical constraints on $f$ and $\theta$, as
    determined by analyzing a simulated \tdv signal measured by {\em
      Kepler}. The signal was based on transits of a Saturn analog on
    a $17.1$-day orbit around a Sun-like star with KIC $r = 13$
    mag. The light gray region marks the 95\% confidence region, using
    a model that allow the parameters $\{ k, f, \theta, i, P_{\rm
      prec}, \phi_0 \}$ to be independent free parameters. In this
    case only the product $f \sin^2 \theta$ is well determined; the
    dashed line satisfies $f \sin^2 \theta = $~constant.  The dark
    gray region is the 95\% confidence region, using a model that
    enforces the expected physical relationship between \tdv amplitude
    and period, as well as the Darwin-Radau relation and a restricted
    range for the normalized moment of inertia $\mathds{C}$.    See
    \S~\ref{sec:case} for details. The dotted lines mark the ``true''
    values of $f$ and $\theta$ that were used to generate the
    simulated \tdv signal.}
   \label{fig:cons}
\end{figure}

For $\di \ne 0$ there are a total of 6 observables listed above, and
it would be possible in principle (with arbitrarily precise data) to
determine all 6 model parameters $k$, $f$, $\theta$, $i$, $P_{\rm
  prec}$ and $\phi_0$. However, using our simulated {\it Kepler} light
curve we find that $f$ and $\theta$ cannot be determined
independently, although Eqn.~(\ref{eq:quant2}) could be used to place
a lower bound on $f$. The degeneracy is illustrated in
Fig.~(\ref{fig:cons}).

To break the parameter degeneracy, one possibility is to arrange for
high-cadence, high-precision observations of at least one transit,
seeking the slight oblateness-induced anomalies that were described by
Seager \& Hui (2002) and Barnes \& Fortney (2003). Observations of a
single light curve would lead to constraints on the sky-projected
oblateness and obliquity [see Eqns.~(\ref{eq:fperp}) and
(\ref{eq:thetaperp})], which, together with the T$\delta$V signal,
would uniquely determine $f$ and $\theta$. The best time to schedule
such observations would be near a minimum of the \tdv curve, when the
light curve anomalies are largest.

This would be a challenging task, as the amplitude of the differences
between the actual light curve and the best-fitting model of the
transit of a spherical planet would be $\lesssim$100~ppm. For this
specific example, even {\em Kepler} photometry (with 1~min cadence)
would be insufficient to detect the anomalies in a single transit. By
observing 10 transits near the minimum of the \tdv~signal (during
which time the sky-projected quantities are constant to within 1\%),
{\it Kepler} could detect the signatures of oblateness and obliquity
at the 1$\sigma$ level, but the resulting constraints would be weaker
than the constraints determined by an analysis of the \tdv signal. A
significantly larger planet, or brighter host star, would be required
for meaningful constraints.

Another possibility is to enforce additional physically-motivated
relationships between parameters. In particular, we have already shown
that the precession period is a function of $k$, $f$, $\theta$,
$\mathds{C}$ and $J_2$ [by combining Eqn.~(\ref{eq:rot}) with
Eqn.~(\ref{eq:prec})]. A further condition can be imposed on $f$,
$\mathds{C}$ and $J_2$, such as the Darwin-Radau approximation for
planets in hydrostatic equilibrium (Murray \& Dermott 2000):
\begin{eqnarray}
  \frac{J_2}{f} = -\frac{3}{10}+\frac{5}{2}\mathds{C}-\frac{15}{8} \mathds{C}^2. \label{eq:dr}
\end{eqnarray}
Following this path, there are 8 model parameters ($k$, $f$, $\theta$,
$i$, $P_{\rm prec}$, $\phi_0$, $\mathds{C}$, $J_2$) with 2
physically-motivated constraints among them. However there are only 5
quantities that are well determined from the photometric data [$P_{\rm
  prec}$, $\phi_0$, $f\sin^2\theta$, $k^2(1-f)$, $i$], leaving us
still short by one observable or constraint from being able to
determine all the parameters. For example, if one were willing to
assume $\mathds{C} = 0.23$, then we find from our simulated {\it
  Kepler} data that $f$, $\theta$, $J_2$ and $P_{\rm prec}$ can be
recovered with a precision of about $10\%$, and $k$ is recovered
within $1\%$.

Assigning $\mathds{C}$ a specific value is unrealistic, but for
realistic planets one expects $\mathds{C}$ to be smaller than 0.4
(Murray \& Dermott 2000). We repeated the analysis of our hypothetical
\tdv signal, allowing $\mathds{C}$ to be a free parameter restricted
to that range, with a uniform prior.\footnote{We also required
  $J_2>0$, which corresponds to $\mathds{C}>0.133$ according to the
  Darwin-Radau relation (Eqn.~\ref{eq:dr}).} In effect we averaged the
results over a range of $\mathds{C}$ deemed to be physically
plausible. As might be expected, a strong degeneracy was observed
between $f$ and $\theta$, as seen in Fig.~(\ref{fig:cons}). However,
we were still able to determine $J_2$ to within $10\%$, and $k$ to
within 2\%.

\section{Discussion}
\label{sec:disc}

In this paper we have investigated the observability of changes to
transit light curves resulting from the spin precession of an oblate,
oblique exoplanet.  The most readily detectable signal is the \tdv
signal, the variation in transit depth due to the changing area of the
planetary silhouette.  The planets that seem most likely to exhibit
detectable effects are those with periods between 15--30 days (around
Sun-like stars), which is short enough for precession periods to be
40~yr or less, and long enough to hope that tidal spin-orbit
synchronization has not taken place.     

It is also important to consider other physical processes that could
give rise to T$\delta$V signals, and which might confound the
interpretation of the data. Starspots and other types of stellar
variability can produce transit depth variations. These can be
recognized and taken into account by monitoring the star outside of
transits, as is done automatically by the {\it Kepler} satellite. We
are not aware of any atmospheric phenomena associated with the
exoplanet that would result in variations in the projected area at the
1\% level, mimicking those due to uniform spin precession.

However, we can think of two plausible phenomena that would affect the
T$\delta$V signal: moons and rings. If the planet has any moons then
the precession period may be shorter than we have calculated.
Satellites provide more leverage for the star to torque the
exoplanetary system.  Mathematically, satellites augment the effective
values of $J_2$ and $\mathds{C}$. Following Ward \& Hamilton (2004) we
may write the enhanced values as $J_2+j$ and $\mathds{C}+c$, where
\begin{eqnarray}
c & = & \sum_i \frac{m_i}{M_p} \left(\frac{a_i}{R_{\rm eq}}\right)^2 \frac{P_{\rm rot}}{p^i_{\rm orb}}\;\;\;~\mbox{and}\\
j & = &\frac{1}{2} \sum_i \frac{m_i}{M_p} \left(\frac{a_i}{R_{\rm eq}}\right)^2 \frac{\sin \left(\theta-I_i\right)}{\sin \theta}
\end{eqnarray}
where $m_i$, $a_i$, $I_i$ and $p^i_{\rm orb}$ are the satellites'
masses, orbital radii, orbital inclinations (relative to the planetary
equator), and orbital periods.

For example, in the Saturnian system $j/J_2 \approx 3.2$ while
$c/\mathds{C} \approx 0.01$ such that $(\mathds{C}+c)/(J_2+j) \approx
3.2$ (as compared to $\mathds{C}/J_2 = 13.5$).  Titan alone is
responsible for about $90\%$ of $j$ and $c$, shortening Saturn's
precession period by a factor of four relative to a satellite-free
Saturn.\footnote{For Titan: $a_i/R_{\rm eq} = 20.2$, $m_i/M_p =
  2.3\times10^{-4}$, $p^i_{\rm orb}/P_{\rm rot}=38.3$; Murray \&
  Dermott (2000).} Fig.~(\ref{fig:times}) shows the effect of Titan
analogs at various distances around the Saturn analog considered in
this paper. The distances are expressed as fractions of the Hill
radius $r_H$. This effect might be used to implicate the presence of
exomoons, if a T$\delta$V signal were observed and found to correspond
to an effective $J_2$ too large to be plausibly attributed to the
planet alone.

Ring systems would increase the amplitude of the signal, while leaving
the period unchanged. Optically thick rings that lie within the
equatorial plane of the planet would increase the fractional variation
of the sky-projected area, as the planet precesses. In contrast to
exomoons, ring systems have little mass, and would not significantly
reduce the precessional period. This is the situation in the Saturnian
system (Ward \& Hamilton 2004). It is not certain that rings could
exist around planets having orbital periods between 15 and 30
days. Rings around planets with orbital periods less than $P_{\rm orb}
\approx15$ days would likely be short-lived as a result of
Poynting-Roberston drag amongst other destructive effects (Barnes \&
Fortney 2004). Also, we would be unlikely to find rings composed of
water ice (e.g.\ Saturn's rings) around planets whose orbits are
interior to the snow line ($< 1$ AU). Nevertheless, rings of other
compositions may exist.

In short, rings and moons would each affect the observed T$\delta$V
signal, and in complementary ways. This may introduce some ambiguity
in the estimation of the shape parameters of the rotating planet, but
may also allow the rings and moons to be detectable, issues that we
leave for future work.

For simplicity we have considered only circular orbits.  Planets on
eccentric orbits will undergo apsidal precession and nodal precession,
which will result in time variable stellar torques on the planet and
consequent modifications to the spin-axis precession. For Saturn or
Jupiter analogs at $P_{\rm orb} \gtrsim 15$ days, the apsidal and
nodal precession periods are $\gtrsim$$10^7$~yr and therefore likely
to be irrelevant (Ragozzine \& Wolf 2009).

In addition, we restricted our attention to the simplest case of
uniform spin precession, but in reality the perturbations from other
bodies may cause the spin axis to perform a more complex ballet. For
example, Mars's spin axis tumbles chaotically (Touma \& Wisdom
1992). Saturn's moon Titan causes a 700~yr modulation of Saturn's spin
precession frequency, due to its inclined orbit. Furthermore, Saturn's
spin axis may be trapped in a resonance with Neptune's orbit, causing
it to librate with an angular amplitude of $\gtrsim$$31^\circ$ as it
circulates about the second Cassini state (Ward \& Hamilton
2004). These effects would be manifested as additional time
dependences (``noise'') in the \tdv signal. The effects are impossible
to forecast for exoplanets, depending as they do on the existence of
other bodies and any resonances that may occur.

\acknowledgements We thank Dan Fabrycky, Darin Ragozzine, and members
of the MIT exoplanet discussion group, for helpful conversations.  We
also thank an anonymous referee for helpful comments on an earlier
draft of this manuscript.

\appendix

\section{The projection of a spheroidal planet onto the plane of the sky}  
   We model the planet as an oblate spheroid, illustrated in
Figure~(\ref{fig:dia}). The sky plane projected shadow of the spheroidal planet is bounded by an ellipse. The polar axis, $\hat{s}$, of the spheroid is tilted by the obliquity angle $\theta > 0$ from the orbital axis $\hat{n}$ and by the angle $\theta' > 0$ from the axis 
$\hat{n}'$ which is perpendicular to the plane that is perpendicular to the sky plane. Let $\hat{y}$ be the axis which lies along the line connecting the center of mass and the observer.  Let $\hat{x}$ be the axis in the orbital plane whose projection into the 
plane perpendicular to the sky plane lies along $\hat{y}$.
 The angle $\phi$ is the angle through which $\hat{s}$ is rotated about $\hat{n}$ and $\phi'$ is the angle through which $\hat{s}$ is rotated about $\hat{n}'$ such that $\phi = \phi' = 0$ corresponds to the orientation in which the spheroid is tipped 
towards the observer (such that $\hat{s}$ is coplanar with both $\hat{x}$ and $\hat{y}$).  It follows that
 \begin{eqnarray}
 	\cos \theta &=& \hat{s}\cdot\hat{n} \\
	\cos \theta' &=& \hat{s}\cdot\hat{n}' \\
	\cos \phi \sin \theta &=& \hat{s}\cdot\hat{x} \\
	\cos \phi' \sin \theta'&=& \hat{s}\cdot\hat{y}\\
	\hat{x}-\hat{y} & = & \hat{n}~\cos i - \hat{x} ~\left(1-\sin i\right)
\end{eqnarray}
where $i \equiv \sin^{-1} \hat{x}\cdot\hat{y}$ is the inclination of the orbital plane to the sky plane and where we have assumed the distance from planet to star is much less than the distance from planet to observer.  The following relationship between unprimed 
angles (measured relative to the orbital plane) and primed angles (measured relative to the sky plane) may be derived from the above equations:
\begin{eqnarray}
	\cos \phi' \sin \theta' & = & \sin i ~\cos \phi \sin \theta+\cos i ~\cos \theta.
\end{eqnarray}
 
The polar axis of the spheroid is inclined relative to the plane of the sky by the angle $\theta''$ where
\begin{eqnarray}
	\cos^2\theta'' & = & 1-\cos^2 \phi' \sin^2 \theta'  \\
			& = & 1-\left( \sin i ~\cos \phi \sin \theta+\cos i ~\cos \theta\right)^2
\end{eqnarray}
If $\hat{p}$ and $\hat{q}$ are the orthonormal axes in the sky plane that are aligned with the axes of the ellipse bounding the planet's shadow, then a point $(y,p,q)$ on the surface of the spheroid satisfies
\begin{eqnarray}
	(y')^2+p^2+\left(\frac{q'}{1-f}\right)^2 = R_{\rm eq}^2 \label{eq:ell}
\end{eqnarray}
where 
\begin{eqnarray}
	y' &=& y \cos \theta''-q \sin \theta''  \nonumber\\
	q' &=& y \sin \theta''+q \cos \theta'',  \nonumber
\end{eqnarray}
$R_{\rm eq}$ is the equatorial radius and $f$ is the oblateness parameter.

The analytic description of the closed curve bounding the projection of the planet on the sky plane may be found by solving for the non-degenerate solutions of $y(p,q)$ in Eqn.~(\ref{eq:ell}).  Non-degenerate solutions correspond to points in the sky plane where rays parallel with $\hat{y}$ intersect at exactly one point on the planetary surface and thus define the boundary of the shadow as seen by the observer (who is located at a distant point along $\hat{y}$).  As advertised, it may be shown that the collection 
of these points satisfy the equation of an ellipse,
\begin{eqnarray}
	p^2+\frac{q^2}{\sin^2 \theta''+(1-f)^2\cos^2 \theta''} &=& R_{\rm eq}^2, \label{eq:boundell}
\end{eqnarray}
having major and minor axis lengths, $A$ and $B$, satisfying
\begin{eqnarray}
	A &=& R_{\rm eq} \\
	B &=& R_{\rm eq}\sqrt{\sin^2 \theta''+(1-f)^2\cos^2 \theta'' }  \nonumber \\
		&=&  R_{\rm eq}\sqrt{1-\epsilon^2\cos^2 \theta''} \nonumber \\
		&=& R_{\rm eq}\sqrt{1-\epsilon^2 \left[1-\left( \sin i ~\cos \phi \sin \theta+\cos i ~\cos \theta\right)^2\right]}
\end{eqnarray}
where $\epsilon \equiv \sqrt{1-(1-f)^2}$.

The minor axis of the ellipse lies along the projection of the polar axis onto the sky plane.
The angle between the major axis of the ellipse and the direction of orbital motion is given as
\begin{eqnarray}
	\theta_\perp = \tan^{-1} \frac{\sin \phi \tan \theta}{\cos \phi \tan \theta \cos i - \sin i}
\end{eqnarray}
and the oblateness parameter of the ellipse, $f_\perp \equiv (A-B)/A$, is given as
\begin{eqnarray}
	f_\perp & = &1- \sqrt{1 - \epsilon^2 \left\{1-\left[\sin \theta\cos\phi \sin i + \cos\theta \cos i\right]^2\right\}}.
\end{eqnarray}
The boundary of the ellipse may be defined relative to the ellipse center via the parameter equations
\begin{eqnarray}
	X(s) & = & R_{\rm eq} \left[ \cos s \cos \theta_\perp-\left(1-f_\perp\right) \sin s \sin \theta_\perp \right] \\
	Y(s) & = & R_{\rm eq} \left[ \cos s \sin \theta_\perp+\left(1-f_\perp\right) \sin s \cos \theta_\perp \right]
\end{eqnarray}
for $s \in [0, 2 \pi]$ and where $X$ is the coordinate along the direction of orbital motion.  The radius, $R_\parallel$, of the projection of the ellipse onto 
the direction of orbital motion is equal to the maximum of $X(s)$.  This maximum occurs for $s_{\rm  max}$ satisfying
\begin{eqnarray}
	\tan s_{\rm max} &=& -(1-f_\perp) \tan \theta_\perp
\end{eqnarray}
for which
\begin{eqnarray}
	R_\parallel \equiv X(s_{\rm max}) & = & R_{\rm eq} \sqrt{\left(1-f_\perp\right)^2 \sin^2 \theta_\perp + \cos^2 \theta_\perp}.
\end{eqnarray}

The areal ratio of the spheroidal planet's sky projection and the stellar disk is
\begin{eqnarray}
	\delta &= &\frac{\pi A B}{\pi R_\star^2} \nonumber \\
		& = & k^2 \sqrt{1-\epsilon^2 \left[1-\left( \sin i ~\cos \phi \sin \theta+\cos i ~\cos \theta\right)^2\right]}.
\end{eqnarray}
where $k \equiv R_{\rm eq}/R_\star$.

\end{document}